\documentclass[12pt,onecolumn,draftclsnofoot]{IEEEtran}
\usepackage[T1]{fontenc}
\usepackage[latin9]{inputenc}
\usepackage{color}
\usepackage{amsmath}
\usepackage{graphicx}

\makeatletter

\providecommand{\tabularnewline}{\\}

\usepackage{caption}
\usepackage{amsmath}

\usepackage{caption}
\usepackage{graphicx}
\usepackage{subfigure}
\usepackage{float}
\usepackage{array}
\usepackage{latexsym}
\usepackage{paralist}
\usepackage{leftidx}
\usepackage{mathrsfs}
\usepackage{multirow}
\usepackage{hyperref}
\usepackage{threeparttable}
\usepackage{pifont}
\usepackage{amsthm}
\IEEEoverridecommandlockouts
\newcommand{\cmark}{\ding{52}}
\makeatother

\begin{document}

\title{Mobility-Aware Caching for Content-Centric Wireless Networks: Modeling
and Methodology}

\author{\IEEEauthorblockN{Rui Wang, Xi Peng, Jun Zhang and Khaled B. Letaief, \textit{Fellow,
IEEE}}\\
\thanks{The authors are with the Department of Electronic and Computer Engineering,
the Hong Kong University of Science and Technology, Clear Water Bay,
Kowloon, Hong Kong (e-mail: \{rwangae, xpengab, eejzhang, eekhaled\}@ust.hk).
Khaled B. Letaief is also with Hamad bin Khalifa University, Doha,
Qatar (e-mail: kletaief@hkbu.edu.qa).}\thanks{This work was supported by the Hong Kong Research Grant Council under
Grant No. 610113.}}
\maketitle
\begin{abstract}
As mobile services are shifting from ``connection-centric'' communications
to ``content-centric'' communications, content-centric wireless
networking emerges as a promising paradigm to evolve the current network
architecture. Caching popular content at the wireless edge, including
base stations (BSs) and user terminals (UTs), provides an effective
approach to alleviate the heavy burden on backhaul links, as well
as lowering delays and deployment costs. In contrast to wired networks,
a unique characteristic of content-centric wireless networks (CCWNs)
is the mobility of mobile users. While it has rarely been considered
by existing works in caching design, user mobility contains various
helpful side information that can be exploited to improve caching
efficiency at both BSs and UTs. In this paper, we present a general
framework on mobility-aware caching in CCWNs. Key properties of user
mobility patterns that are useful for content caching will be firstly
identified, and then different design methodologies for mobility-aware
caching will be proposed. Moreover, two design examples will be provided
to illustrate the proposed framework in details, and interesting future
research directions will be identified.

\newpage{}
\end{abstract}

\section{Introduction }

\textcolor{black}{Mobile data traffic is undergoing an unprecedented
growth, and it is being further propelled by the proliferation of
smart mobile devices, e.g., smart phones and tablets. In particular,
the data services subscribed by mobile users have gradually shifted
from \textquotedblleft connection-centric\textquotedblright{} communications,
e.g., phone calls and text messages, to \textquotedblleft content-centric\textquotedblright{}
communications, e.g., multimedia file sharing and video streaming.
One main effort to meet such a strong demand is to boost the network
capacity via network densification, i.e., to deploy more access points.
While this approach is expected to significantly increase the capacity
in future 5G networks, it incurs a tremendous demand for backhaul
links that connect the access points to the backbone network. Thus,
it will cause a heavy financial burden for mobile operators who are
required to upgrade the backhaul network, and such a comprehensive
approach will not be cost-effective to handle content-centric mobile
traffic, which may be bursty and regional. Consequently, a holistic
approach is needed and }\textit{\textcolor{black}{cache-enabled content-centric
wireless networking}}\textcolor{black}{{} emerges as an ideal solution.}

\textcolor{black}{Nowadays, abundant caching storages are available
at the }\textcolor{black}{\emph{wireless edge}}\textcolor{black}{,
including both base stations (BSs) and user terminals (UTs), which
can be used to store popular contents that will be repeatedly requested
by users. Since the prices of caching devices, e.g., solid state drives
(SSDs), have been coming down year after year, it has become more
and more cost-effective to deploy caches instead of laying high-capacity
backhaul links \cite{cachebenefit}. Moreover, the ample storages
at mobile UTs, currently as large as hundreds of gigabytes, are also
potential resources to be utilized for caching. Besides reducing the
demand and deployment costs of backhaul links, caching popular content
is also an effective technique to lower delays and reduce network
congestion \cite{cachebenefit2}, since mobile users may acquire the
required files from the serving BSs or the proximal UTs directly without
connecting to the backbone network.}

\textcolor{black}{The idea of content-centric networking has already
been explored in wired networks, where named pieces of content are
directly routed and delivered at the packet level, and content packets
are automatically cached at routers along the delivery path. Accordingly,
caching design at the routers, including content placement and update,
is crucial to the system performance. Caching at the wireless edge
can draw lessons from its wired counterpart, but it also enjoys new
features. The broadcast nature of the radio propagation will fundamentally
affect the content caching and file delivery, which has recently attracted
significant attention. Another important feature of content-centric
wireless networks (CCWNs) is user mobility, which has been less well
studied. While mobility imposes additional difficulties on caching
design in CCWNs, it also brings about new opportunities. User mobility
has been proved to be a useful feature for wireless network design,
e.g., it has been utilized to improve the routing protocol in wireless
ad hoc networks \cite{exintercontactmodel}. Unfortunately, most previous
studies on caching design in CCWNs ignored user mobility and assumed
fixed network topologies, which cannot capture the actual scenario.
There have been initial efforts on caching designs by incorporating
user mobility }\cite{femtomobility}\textcolor{black}{. However, only
some special properties of user mobility patterns were addressed and
there is a lack of systematic investigation.}

The main objective of this paper is to provide a systematic framework
that can take advantage of user mobility to improve the caching efficiency
in CCWNs. Specifically, a comprehensive discussion of spatial and
temporal properties of user mobility patterns will firstly be provided,
each of which will be linked to specific caching design problems.
We will then propose mobility-aware caching strategies, with two typical
design cases as examples. Finally, we will identify some future research
directions.

\begin{figure}[!t]
\centering \includegraphics[width=4.5in]{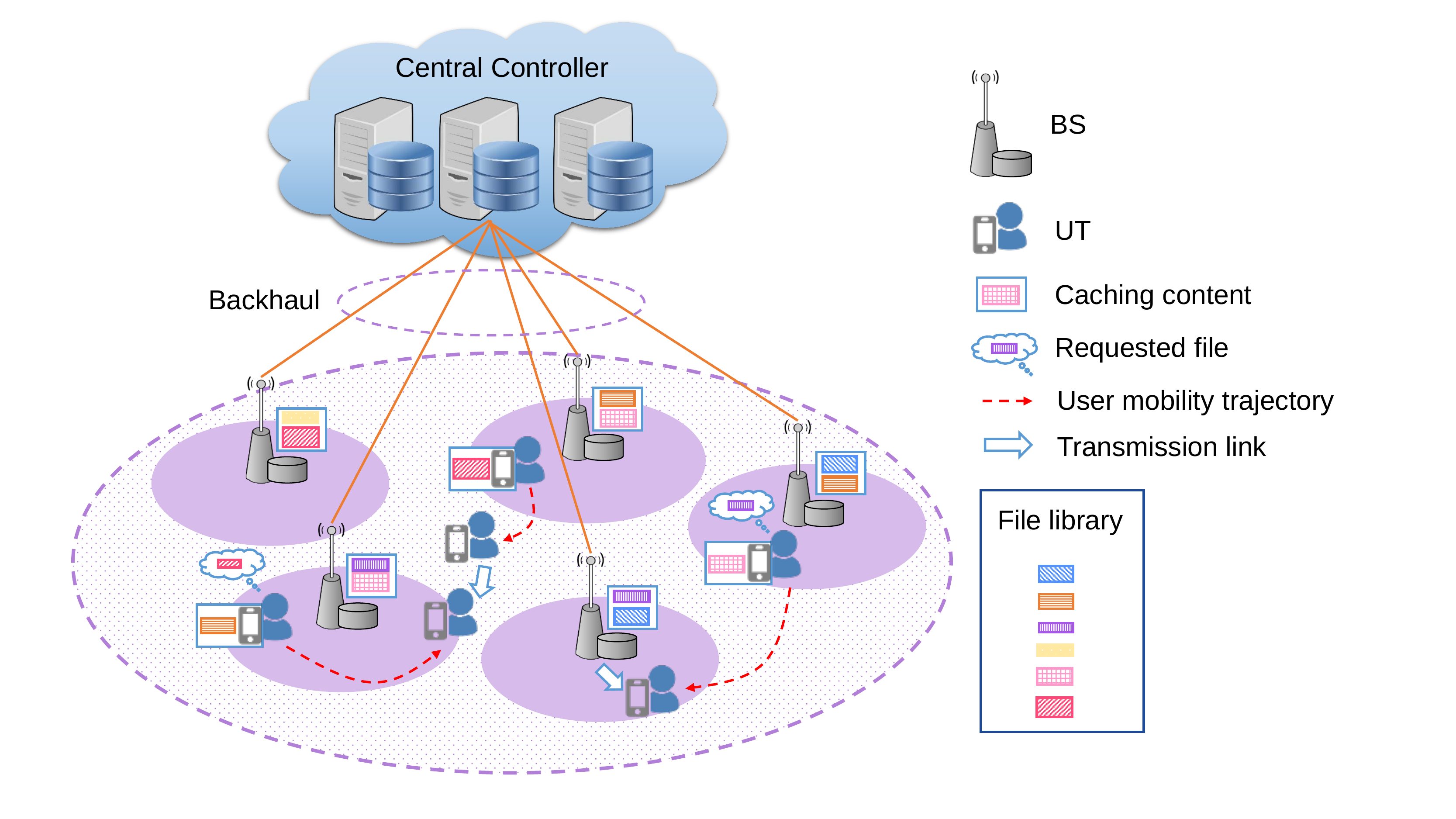} \caption{A sample \textcolor{black}{cache-enabled CCWN. }A mobile user may
download the requested file from the BSs or UTs along its moving path
that have this file in cache. Once the requested files match the cached
data, transmissions over the backhaul network will be avoided. Otherwise,
mobile users have to request from the central controller via backhaul
links.}
\label{model} 
\end{figure}

\section{Exploiting User Mobility in Cache-Enabled CCWNs }

\textcolor{black}{In this section, we will illustrate the importance
of considering user mobility when designing caching strategies in
CCWNs. A sample cache-enabled CCWN is shown in Fig. \ref{model},
where both BSs and UTs have cache storages and are able to cache some
pieces of content from the file library. In the following, we will
first introduce the main caching design problems in CCWNs, and then
identify important properties of the user mobility patterns and associate
them with different caching problems.}

\subsection{\textcolor{black}{Key Design Problems of Caching in CCWNs}}

\textcolor{black}{The fundamental problem in caching design for CCWNs
is to determine where and what to cache. The design principles may
depend on different types of side information, including }\textit{\textcolor{black}{long-term
information}}\textcolor{black}{{} obtained from observations over a
long period of time, such as the statistics of users' requests and
average communication times with BSs and other UTs, and }\textit{\textcolor{black}{short-term
information}}\textcolor{black}{{} generated by instant changes, e.g.,
instantaneous channel state information and real-time location information.
The collection of long-term information incurs a low overhead, while
the usage of short-term information can provide better performance
but requires frequent update. In the following, we categorize different
caching design problems in CCWNs according to the timeliness of the
available information.}

\subsubsection{\textcolor{black}{Caching Content Placement}}

\textcolor{black}{Caching content placement typically relies on long-term
system information and is used to determine how to effectively pre-cache
content in the available storage. To reduce overhead, the update of
side information and caching content will not be very frequent. It
is normally assumed that the long-term file popularity distribution
is known as a priori, and the network topology can either be fixed
or subject to some assumptions in order to simplify the design.}

\textcolor{black}{Previous works have provided some insights into
caching content placement at BSs. In particular, without cooperation
among BSs, the optimal caching strategy is to store the most popular
files \cite{cache_coding_Placement}. However, if users are able to
access several BSs, each user will see a different but correlated
aggregate cache, and in this scenario, allocating files to different
BSs becomes nontrivial. Moreover, the coded caching scheme, where
segments of Fountain-encoded versions of the original file are cached
\cite{cache_coding_Placement}, outperforms the uncoded caching scheme
where only complete files are cached. By carefully designing the caching
content placement via combining multiple files with a given logic
operator, different requests can be served by a single multicast transmission
\cite{Codedcaching}, which results in a significant performance improvement
compared to the uncoded scheme. }

\textcolor{black}{Meanwhile, caching content placement at UTs is also
attracting noticeable attention. Caching at UTs may allow users to
download requested content in a more efficient way with device-to-device
(D2D) communications, where proximal users communicate with each other
directly. Compared with caching at BSs, the advantages of caching
at UTs come from the lower deployment costs and an automatic promotion
of the storage capacity when the UT density increases, as the ensemble
of UTs forms an aggregate cache; while the drawbacks include the difficulty
of motivating UTs to join the aggregate cache, and the more complicated
randomness in the D2D scenario. Pioneering works have shed light on
caching content placement at UTs \cite{D2D}.}

\textcolor{black}{However, it is noted that previous studies rarely
considered user mobility, which can be tracked without much difficulty
with today's technologies. If we could make use of long-term statistics
of user mobility, such as the average steady-state probability distribution
over BSs, the efficiency of content caching will be significantly
improved.}

\subsubsection{Caching Content Update}

Though long-term information incurs a low overhead to obtain, it contains
less fine grained information, which may also expire after a period
of time and thus cannot assure accuracy. For example, the BS-UT or
UT-UT connectivity topology may change quickly due to the movement
of UTs. Consequently, it may cause significant errors by using the
expired long-term information to design caching strategies. If short-term
information is available, such as the real-time information of the
file requests and transmission links, caching content can be updated
to provide a better experience for mobile users. In the following,
we will introduce two caching content update problems.

\paragraph{Adaptive caching}

Since caching storage is limited, it is critical to replace the stale
caching content to improve caching efficiency. Common adaptive caching
schemes to increase the cache hit ratio include replacing the least
recently used content and replacing the least likely requested content
\cite{videoAmazon}. Another typical application of adaptive caching
is to serve the users that follow regular mobility patterns and have
highly predicable requirements. When the mobility regularity and request
preference of mobile users are known, BSs can update the caching content
according to the estimation of future requests. The main challenges
come from the accurate prediction of users' future positions and requirements,
the frequency to conduct the adaptive caching strategy, as well as
the replacement priorities for the caching content.

\paragraph{Proactive caching}

In practice, a user can only download a portion of its requested file
rather than the entire file from a BS, as the moving user may not
have enough communication time with the BS. Proactive caching aims
at providing seamless handover and downloading for users by pre-fetching
the requested content at the BSs that will be along the users' future
paths with a high probability. Nevertheless, user requests and locations
are usually unknown in realistic environments, and thus the accuracy
of location prediction is critical to the performance.

\subsection{Modeling User Mobility Patterns}

\begin{figure}
\centering \subfigure[User trajectory]{ %
\begin{minipage}[b]{0.45\textwidth}%
 \includegraphics[width=1\textwidth]{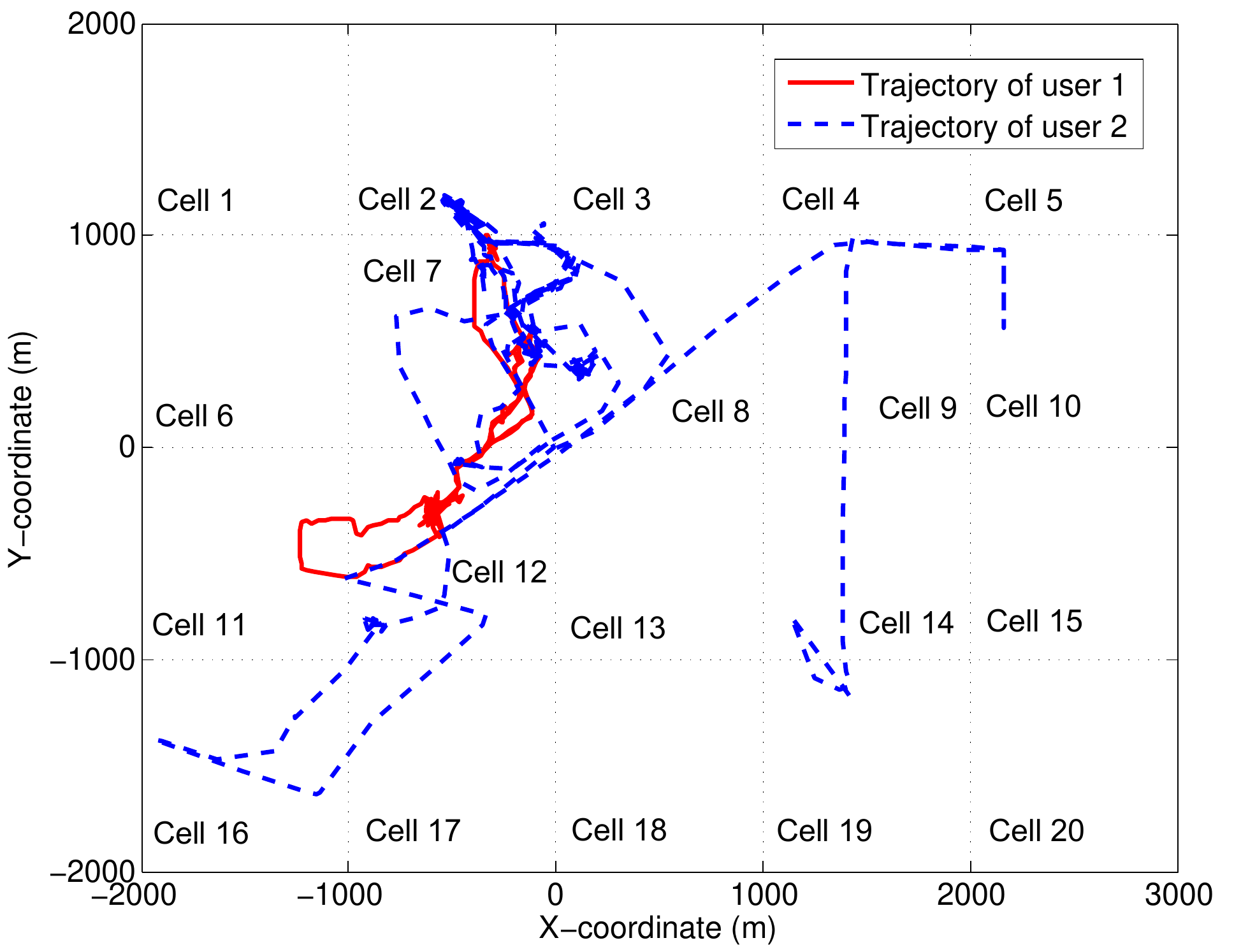} %
\end{minipage}\label{trace} } %
\begin{minipage}[b]{0.45\textwidth}%
 \subfigure[Cell sojourn times]{ \includegraphics[width=1\textwidth]{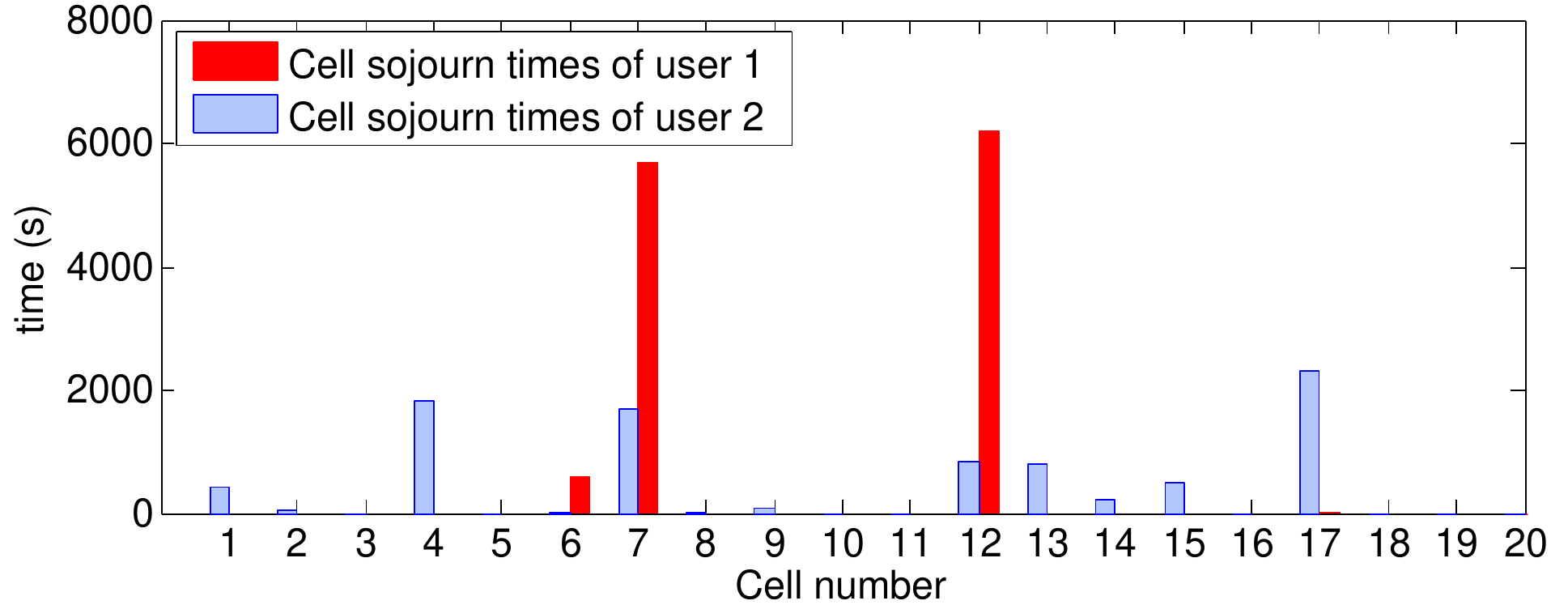}
\label{sojourn} } \\
 \subfigure[Timeline of two users]{ \includegraphics[width=1\textwidth]{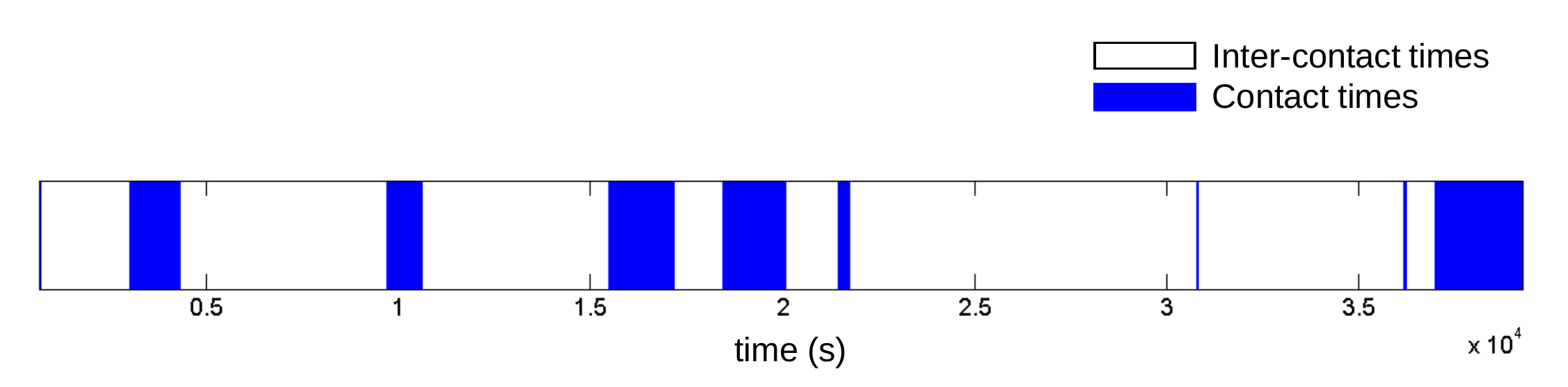}
\label{intercontact} } %
\end{minipage}\caption{The trajectories of two mobile users based on data collected on a
university campus. The two users are moving within a $5000\text{ {m}}\times4000\text{ {m}}$
area. We assume that $20$ BSs are deployed regularly in the area,
with the cell indices labeled in (a). The average cell sojourn times
of these two users, which denote the duration of the users being connected
to each BS, are shown in (b). The transmission ranges of two mobile
users are assumed to be $200$ m and the timeline of users 1 and 2,
including inter-contact times and contact times, is depicted in (c).}
\end{figure}

As can be inferred from the above discussions, taking user mobility
into consideration is critical for the caching design in CCWNs. In
this subsection, we will provide detailed descriptions of different
user mobility properties, which can be classified into two categories,
i.e., the spatial and temporal properties. The spatial properties
contain the information of user mobility patterns related to the physical
locations, while the temporal properties characterize the time-related
features.

\subsubsection{Spatial Properties}

The mobility pattern of a mobile user can be visualized by the \emph{user
trajectory}, i.e., the user's moving path. Crucial information for
caching design in CCWNs, e.g., serving BSs, and distances between
BSs and mobile users, can be obtained from the trajectories of the
mobile users. It is an ongoing research topic to investigate realistic
models for user trajectory, e.g., the random waypoint model in \cite{randomwaypoint}.
As an example, the trajectories of two mobile users are shown in Fig.
\ref{trace}, which are based on data collected on a university campus
\footnote{I. Rhee, M. Shin, S. Hong, K. Lee, S. Kim, and S. Chong, \textquotedblleft CRAWDAD
dataset ncsu/mobilitymodels (v. 2009-07-23),\textquotedblright{} Downloaded
from http://crawdad.org/ncsu/mobilitymodels/20090723, Jul. 2009.}.

The \emph{cell transition}, which denotes the transition pattern of
a user moving from one cell to another, implies the information of
serving BSs for each mobile user, which is one of the most critical
pieces of information in caching design at BSs. Compared to the user
trajectory, the cell transition contains less fine grained information
as the moving path inside each cell cannot be specified. It is appropriate
to capture the transition property using a Markov chain model \cite{MVchain},
where the number of states equals the number of BSs. In the Markov
chain, one state denotes a specific user being served by a given BS,
and the transition probabilities represent the probabilities for a
specific user moving from the serving area of one BS to that of another
BS.

Recently, it has been found that user mobility patterns also largely
depend on the social relations among mobile users. For example, it
was claimed in \cite{sociallink} that the mobile users having relatively
strong social ties are more likely to have similar trajectories. In
\cite{musolesi2004ad}, Musolesi \emph{et al.} proposed a two-level
mobility model, which first establishes a social graph, where the
nodes represent mobile users and the weighted edges represent the
strength of the social connection between mobile users. Then, \textit{social
groups} are built and mobile users in each group move together. Such
information will be useful for caching at UTs.

\subsubsection{Temporal Properties}

To capture the information of the frequency and duration that two
mobile users are connected with each other, the timeline of an arbitrary
pair of mobile users is represented by \emph{contact times} and \emph{inter-contact
times}, where the \emph{contact times} are defined as the time intervals
during which the mobile users are within the transmission range, and
the \emph{inter-contact times} are defined as the time intervals between
two consecutive contact times. The timeline of two users shown in
Fig. \ref{trace} is illustrated in Fig. \ref{intercontact}. Such
a mobility model has been applied to routing problems in ad hoc networks.
For instance, in \cite{exintercontactmodel}, Conan \emph{et al.}
modeled locations of contact times in the timeline of each pair of
mobile users as a Poisson Process so as to capture the average pairwise
inter-contact times in an ad hoc network.

The \textit{cell }\emph{sojourn time }denotes the time duration of
a specific user served by a given BS, which may affect the amount
of data that this user can receive from the BS. Fig. \ref{sojourn}
shows the cell sojourn times of the two users whose trajectories are
shown in Fig. \ref{trace}. Specifically, in \cite{MVchain}, Lee
\emph{et al.} provided an approach to obtain the sojourn time distributions
according to the associated moving history of mobile users.

The user mobility pattern always possesses a periodic property, which
can be exploited to tackle the caching update problem. The \emph{return
time}, which is defined as the time for an arbitrary mobile user to
return to a previous visited location, is considered as a measure
to reflect the periodic property and the frequency of mobile users
to revisit a given area. In \cite{returntime}, Gonzales \emph{et
al.} measured the distribution of the return time and figured out
that the peaks of the return time probability are at $24$ h, $48$
h and $72$ h.

\subsection{Exploiting Mobility for Caching in CCWNs}

Built upon the information given in the above two subsections, potential
approaches will now be proposed to take advantage of user mobility
patterns to resolve different caching design problems in CCWNs, as
summarized in Table \ref{tab_mobility}.

\subsubsection{Caching content placement at BSs}

In CCWNs, as a user moves along a particular path, the user may download
the requested file from all the BSs along this path, and different
BSs may cooperatively cache this file to improve the efficiency. For
this purpose, the statistic and predictive information of the BSs
along the user trajectory, which can be obtained based on user trajectory
or cell transition probabilities, will be needed. Compared to cell
transition probabilities, the user trajectory provides additional
information, i.e., different transmission distances from BSs in different
cells, which can help better design the BS cooperative caching in
CCWNs. For example, different transmission distances may result in
different transmission rates, which will affect the amount of data
that can be downloaded from different BSs. Furthermore, the cell sojourn
time is also a critical factor to determine the amount of data that
can be delivered, and thus will also affect the caching content placement
at BSs.

\subsubsection{Caching content placement at UTs}

By enabling caching at UTs, mobile users may get the requested files
via proximal D2D links. For caching design in such a setting, the
information related to inter-user contacts is essential. In particular,
inter-contact times and contact times will be valuable information,
which will be further illustrated in the design examples in the next
section. In addition, social relations may help to decompose a large
network into several small social groups, and thus reduce the complexity
of caching design. Meanwhile, social groups also imply some contact
information, i.e., mobile users in the same social group are more
likely to have more contacts \cite{social-allocation}. Thus, social
group information can also be utilized to design caching content placement
at UTs.

\subsubsection{Adaptive caching}

The caching content can be adjusted adaptively based on the periodical
mobility pattern, for which the knowledge of return times will be
very useful. Moreover, mobile users in different social groups may
have different content preferences. Thus, the mobility pattern of
each social group can be utilized to improve the adaptive caching
design. For example, in a restaurant, there may be several customer
groups with different content preferences during different time periods,
e.g., elders may enjoy the morning tea, students will have lunch with
friends, and office workers may have dinner together. The BSs around
the restaurant may perform adaptive caching updates accordingly.

\subsubsection{Proactive caching}

If the user trajectory or cell transition property can be estimated
based on past data, the future serving BSs for mobile users can be
predicted. In this way, if a mobile user requests a certain file,
the BSs that are predicted to be on its future path may proactively
cache the requested file, each with a certain segment, and then the
user can download the file when passing by. While it may slightly
increase the backhaul traffic, such proactive caching can significantly
improve the caching efficiency and reduce download latency.

The above proposals are by no means complete. Nevertheless, they clearly
indicate the great potential and importance of mobility-aware caching
in CCWNs. We hope this discussion will inspire more follow-up investigations.

\begin{table}
\caption{Exploiting Mobility for Caching in CCWNs}
\label{tab_mobility} \centering %
\begin{tabular}{|p{1in}<{\centering}|p{0.69in}<{\centering}|p{0.69in}<{\centering}|p{0.69in}<{\centering}|p{0.69in}<{\centering}|p{0.69in}<{\centering}|p{0.69in}<{\centering}|}
\hline 
 & \multicolumn{3}{c|}{Spatial Properties} & \multicolumn{3}{c|}{Temporal Properties}\tabularnewline
\cline{2-7} 
 & User trajectory  & Cell transition  & Social group  & User inter-contact time  & Cell sojourn time  & Return time \tabularnewline
\hline 
Caching content placement at BSs  & \cmark  & \cmark  & \textbf{---}  & \textbf{---}  & \cmark  & \textbf{---} \tabularnewline
\hline 
Caching content placement at UTs  & \textbf{---}  & \textbf{---}  & \cmark  & \cmark  & \textbf{---}  & \textbf{---} \tabularnewline
\hline 
Adaptive caching  & \textbf{---}  & \textbf{---}  & \cmark  & \textbf{---}  & \textbf{---}  & \cmark \tabularnewline
\hline 
Proactive caching  & \cmark  & \cmark  & \textbf{---}  & \textbf{---}  & \textbf{---}  & \textbf{---} \tabularnewline
\hline 
\end{tabular}

\begin{tablenotes} 

\item General: '\cmark ' means that the mobility property can be
utilized in the corresponding caching design problem, and '\textbf{---}'
means that the mobility property may not be utilized. \end{tablenotes} 
\end{table}

\section{Mobility-Aware Caching Content Placement}

In this section, we present two specific design examples for mobility-aware
caching content placement, including caching at BSs and caching at
UTs. Sample numerical results will be provided to validate the effectiveness
of utilizing user mobility patterns in wireless caching design problems.

\begin{figure}
\centering \subfigure[Caching at BSs]{ %
\begin{minipage}[b]{0.5\textwidth}%
 \includegraphics[width=1\textwidth]{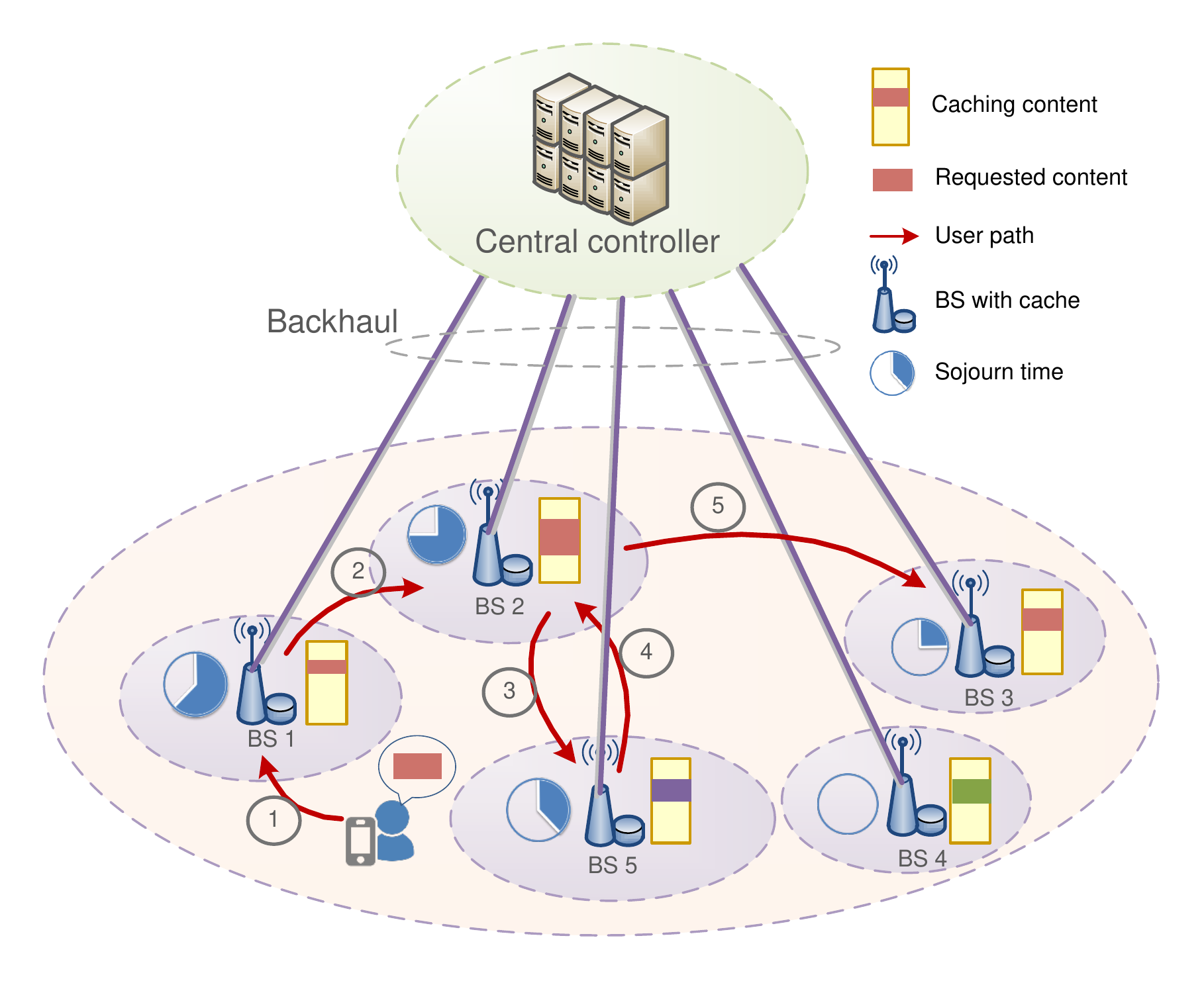} %
\end{minipage}\label{trace-1} }\centering \subfigure[Caching at UTs]{ %
\begin{minipage}[b]{0.45\textwidth}%
 \includegraphics[width=1\textwidth]{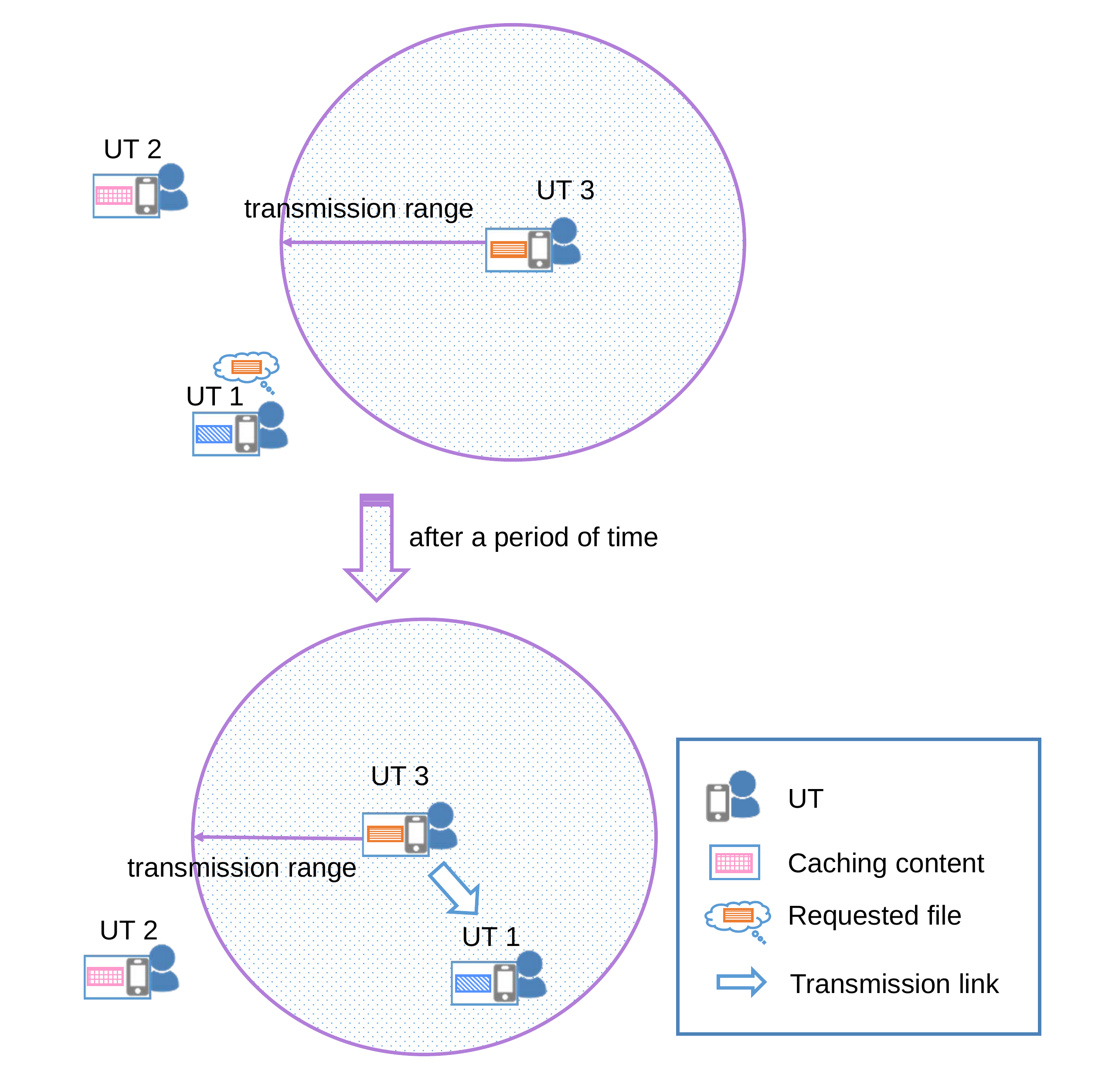} %
\end{minipage}\label{trace-1-1} } \caption{Wireless Caching Networks. BS caching is shown in (a), where a user
requests a file and passes by BSs numbered \{1, 2, 5, 2, 3\} in sequence.
The user can obtain the requested file by collecting data from these
BSs. D2D caching is shown in (b), where UT 1 requests a file, and
it has not stored the file in its own cache. After a period of time,
UT 1 encounters UT 3 which stores the requested file, and it downloads
the file from UT 3.}
\end{figure}

\subsection{Mobility-Aware Caching at BSs}

\textcolor{black}{We first consider utilizing the cell sojourn time
information to design caching content placement at BSs, which may
be macro BSs or femto-cell BSs. A sample network is shown in Fig.
\ref{trace-1}. For simplicity, we assume the downlink rate for each
user is the same while passing by each BS, and cell sojourn times
are estimated based on available data. Mobile users will request files
in the file library based on their demands, which is assumed to follow
a Zipf distribution. Both uncoded and coded caching schemes are considered.
In the uncoded case, we assume that each file is either fully stored
or not stored at each BS. In the coded case, rateless fountain codes
are applied, where each BS may store part of a coded file, and the
whole file can be recovered by collecting enough coded message of
that file \cite{cache_coding_Placement}. When a mobile user requests
a file, the user will try to collect the requested file while passing
by each BS. The proportion of the requested file that can be downloaded
from a BS is limited by the transmission rate and the sojourn time
in this cell, as well as the proportion of the requested file stored
at this BS. We aim to minimize the cache failure probability, which
is the probability that the mobile users cannot get the requested
files from cached contents at BSs. The coded caching placement problem
can be formulated as a convex optimization problem, while the uncoded
caching placement can be obtained by solving a mixed integer programming
(MIP) problem.}

\begin{figure}[!t]
\centering \includegraphics[width=3.4in]{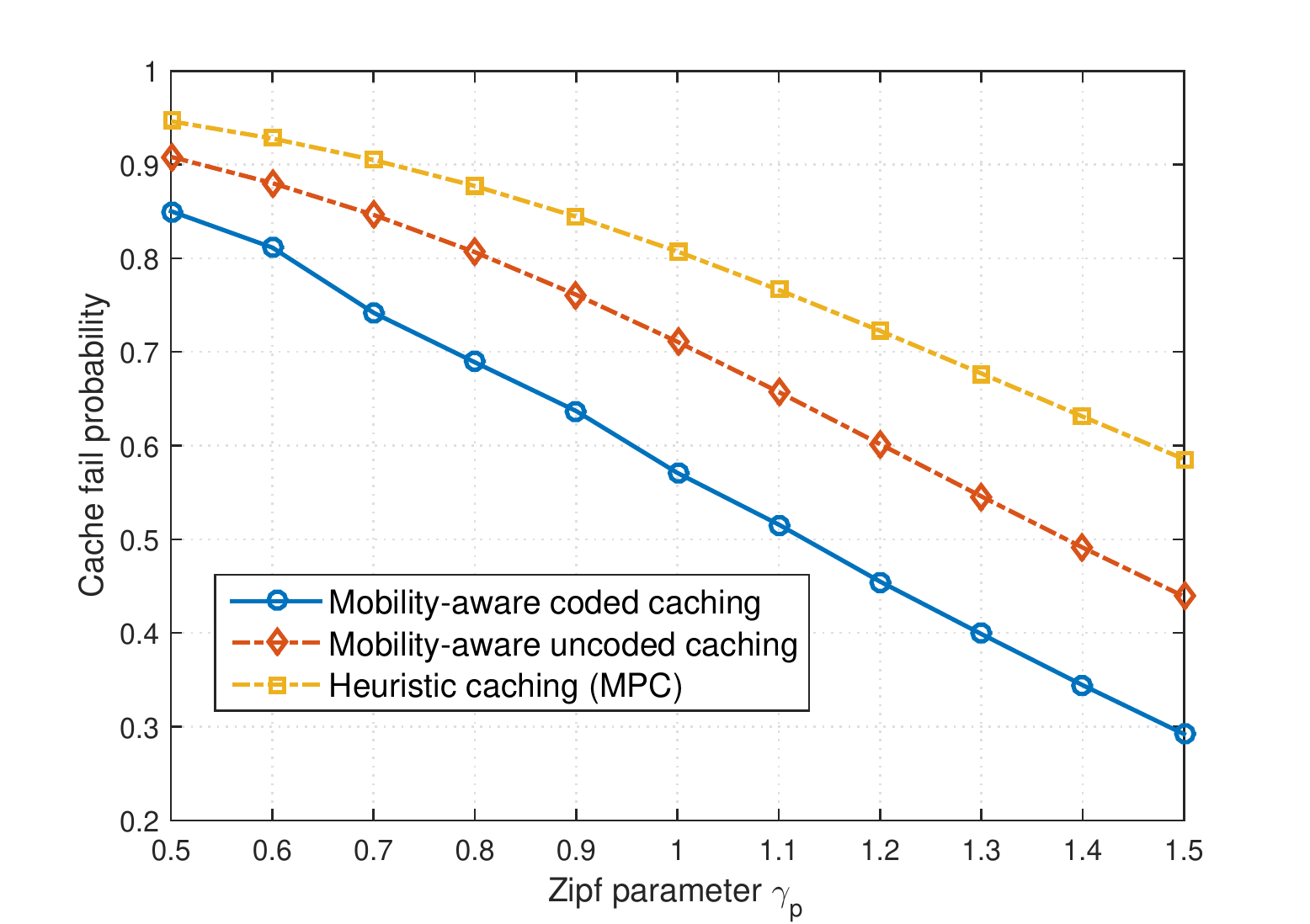} \caption{Comparison of different BS caching content placement strategies, with
6 BSs and a file library of 100 files, while the caching storage capacity
of each BS is the size of one file.}
\label{BScachingdesign} 
\end{figure}

\textcolor{black}{We evaluate the performance of the proposed mobility-aware
caching strategies based on a real-life data set of user mobility,
which was obtained from the wireless network at Dartmouth College
}\footnote{\textcolor{black}{D. Kotz, T. Henderson, I. Abyzov, and J. Yeo, \textquotedblleft CRAWDAD
dataset dartmouth/campus (v. 2009-09-09),\textquotedblright{} Downloaded
from http://crawdad.org/dartmouth/campus/20090909, Sept. 2009.}}\textcolor{black}{. Following caching placement strategies are compared: }
\begin{itemize}
\item \textcolor{black}{{Mobility-aware coded caching strategy}, which
is the proposed coded caching strategy obtained by solving a convex
optimization problem.}
\item \textcolor{black}{{Mobility-aware uncoded caching strategy}, which
is the proposed uncoded caching strategy obtained by solving an MIP
problem.}
\item \textcolor{black}{{MPC strategy}, which is a heuristic caching strategy,
for which each BS stores the most popular contents \cite{videoAmazon}. }
\end{itemize}
\textcolor{black}{The comparison is shown in Fig. \ref{BScachingdesign},
where a larger value of the Zipf parameter $\gamma_{p}$ implies the
requests from mobile users are more concentrated on the popular files.
We see that the mobility-aware caching strategies outperform the heuristic
caching strategy, and the performance gap expands with $\gamma_{p}$,
which demonstrates the value of the mobility information. Moreover,
the coded caching strategy performs better than the uncoded caching
strategy, which validates the advantage of coded caching.}

There are many interesting problems for further investigation. For
example, the user trajectory can be utilized to consider variant download
rates, which will affect the amount of data obtained in different
cells. In addition, based on the user trajectory, it is possible to
jointly deal with the caching problem and interference management.
Another challenge is that many BS caching problems are typically NP-hard,
and thus time-efficient sub-optimal algorithms are needed.

\subsection{Mobility-Aware Caching at UTs}

In this subsection, we will focus on caching at UTs. We consider taking
advantage of average inter-contact times among mobile users to improve
the caching efficiency at UTs. An illustrative example is shown in
Fig. \ref{trace-1-1}.\textcolor{blue}{{} }\textcolor{black}{The locations
of contact times in the timeline for any two mobile users are modeled
as a Poisson process, as in \cite{exintercontactmodel}, where the
intensity is estimated from the history data. For simplicity, the
timelines for different pairs of mobile users are assumed to be independent,
and each file is assumed to be either completely stored or not stored
at each UT. Mobile users will request files in the file library based
on their demands, which is assumed to follow a Zipf distribution.}
When a mobile user generates a request, it will first try to find
the requested file in its own cache, and will then wait for encountering
users storing the requested file. The \emph{delay time} is defined
as the time between when a user requests a file and when it encounters
the first user storing the requested file. We assume that if the mobile
user stores the requested file or its delay time is within a pre-determined
delay threshold, it will be served via D2D links; otherwise, it will
get the file from the BS. To offload the traffic from BSs and encourage
proximal D2D transmissions, we set the objective as to maximize the
\textit{\textcolor{black}{data offloading ratio,}}\textcolor{black}{{}
which is the fraction of users that can get requested files via D2D
links. }This turns out to be a challenging problem and falls in the
category of monotone submodular maximization over a matroid constraint,
which can be solved by a greedy algorithm with an approximation ratio
as $\frac{1}{2}$. 

\begin{figure}[!t]
\centering \includegraphics[width=3.4in]{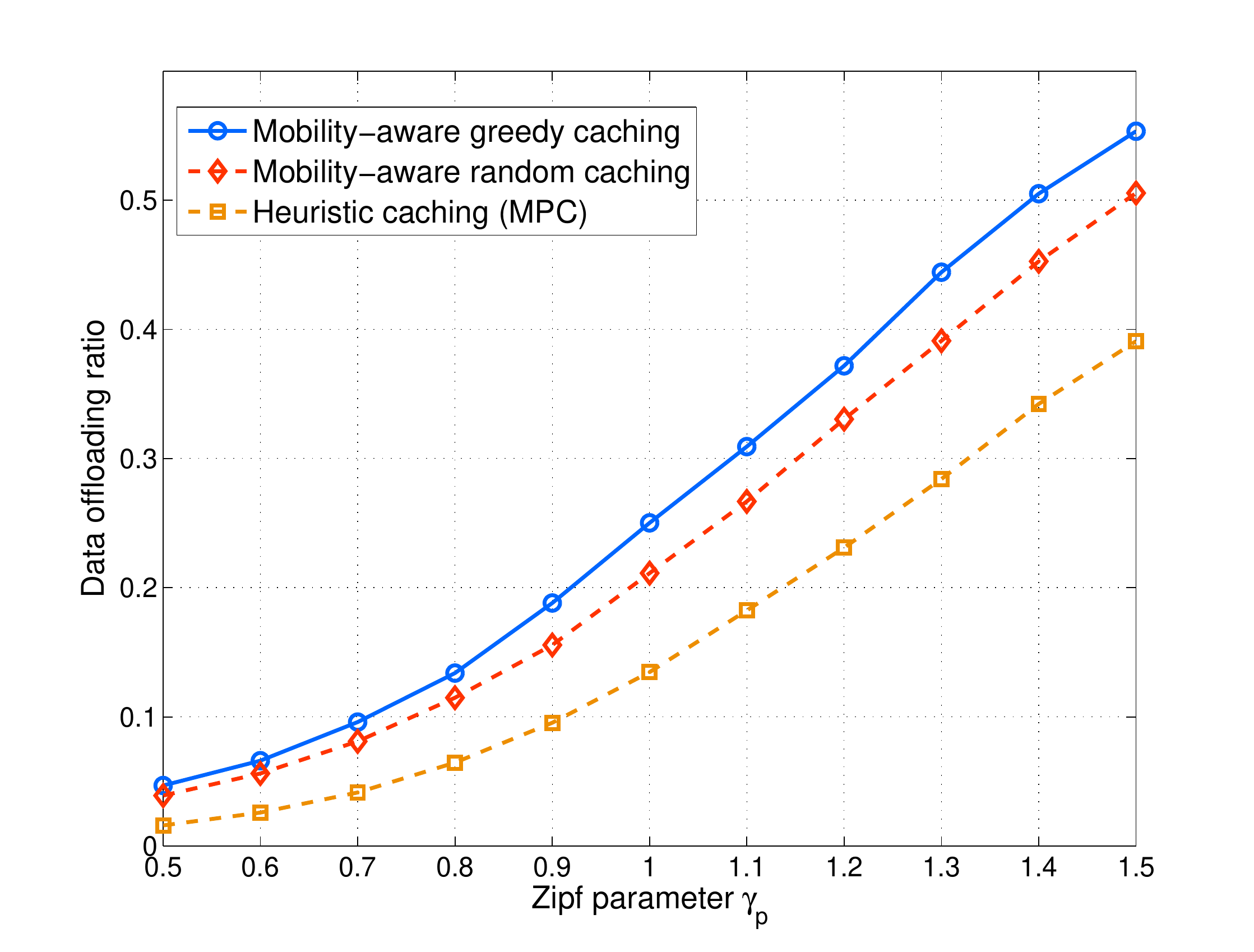} \caption{Comparison of different caching content placement strategies at UTs
with 78 mobile users and a file library consisting of 1000 files,
while each UT can cache at most one file.}
\label{D2Dcachingdesign} 
\end{figure}

The performance of mobility-aware caching at UTs is evaluated based
on a real-life data set, which was collected at the INFOCOM conference\footnote{J. Scott, R. Gass, J. Crowcroft, P. Hui, C. Diot, and A. Chaintreau,
CRAWDAD dataset cambridge/haggle (v. 2009\nobreakdash-05\nobreakdash-29),
downloaded from http://crawdad.org/cambridge/haggle/20090529, doi:10.15783/C70011,
May 2009. } \cite{reallifedata}. Considering that most requests may occur in
the daytime, we generate average inter-contact times according to
the daytime data during the first day of the conference. The following
caching placement strategies are compared: 
\begin{itemize}
\item {Mobility-aware greedy caching strategy}, which is the proposed
caching strategy using a greedy algorithm. 
\item {Mobility-aware random caching strategy}, which is similar to the
random caching strategy proposed in \cite{D2D}. In this strategy,
each UT caches files according to a Zipf distribution with parameter
$\gamma_{c}$. The optimal value of $\gamma_{c}$, which maximizes
the expected fraction of users that can get requested files via D2D
links, is obtained by a line search.
\item {MPC strategy}, which is the same as the one used in Fig. \ref{BScachingdesign}.
\end{itemize}
Based on the data during the daytime in the second day of the conference,
the performance of three caching strategies are compared in Fig. \ref{D2Dcachingdesign}
by varying the file request parameter. \textcolor{black}{It shows
that both mobility-aware caching strategies significantly outperform
the MPC strategy, and the performance gain increases as $\gamma_{c}$
increases. Furthermore, the mobility-aware greedy caching strategy
has a better performance than the mobility-aware random caching strategy,
since the former strategy incorporates average pairwise inter-contact
times more explicitly and allows more optimization variables. Through
extensive simulations, we also observe that, as the number of users
increases, the data offloading ratio using mobility-aware caching
strategies increases, while the MPC strategy remains the same. Meanwhile,
using mobility-aware strategies, the data offloading ratio increases
as the user mobility increases, and the greedy caching strategy always
outperforms the random one. }This implies that a better utilization
of user mobility patterns can further improve the caching efficiency. 

While this initial study provides promising results, lots of challenges
remain. For example, since the number of mobile users in a CCWN is
usually very large, collecting the pairwise inter-contact times will
cause a high overhead. One potential solution is to decompose the
large number of mobile users into several social groups, and then
design caching content placement at UTs based on the inter-contact
times of mobile users within the same social group. Moreover, coded
caching strategies can also be applied, which is a prominent approach
to further optimize the caching efficiency.

\section{Conclusions and Future Directions}

In this paper, we conducted a systematic study that investigated the
exploitation of user mobility information in cache-enabled CCWNs.
Useful spatial and temporal mobility properties were identified and
linked to key caching design problems. Through two design examples,
the advantages and effectiveness of mobility-aware caching were demonstrated.
To fully exploit mobility information in CCWNs, more works will be
needed, and the followings are some potential future research directions.
\begin{itemize}
\item \textit{Joint caching content placement at the wireless edge}: In
practice, many caching systems consist of more than one layer of caches,
which leads to a more complicated hierarchical caching architecture.
In CCWNs, while most existing works, as well as our discussion in
this paper, treated caching at BSs and UTs as separate problems, a
joint design of caching at both BSs and UTs will be essential to further
improve the system performance. 
\item \textit{Dynamic user caching capacities}: Unlike BSs, the caching
capacities at UTs may not be fixed, since they are related to storage
usages of mobile users, which are different from user to user and
are changing over time. It is thus important to investigate how to
adaptively cache according to the dynamic user caching capacities,
while also taking user mobility into consideration. 
\item \textit{Big data analytics for mobility information extraction}: With
the explosive growth of mobile devices, collecting user mobility information
will generate huge amounts of data. Thus, big data analytics to extract
the required mobility information is another challenge in mobility-aware
caching. Meanwhile, accurate prediction is also critical. Though some
existing user mobility models can predict the future mobility behavior
via historical data, e.g., the Markov chain model in \cite{MVchain}
can jointly predict the cell transition and cell sojourn time, more
works will be needed, e.g., on how to predict the user trajectory.
\textcolor{black}{It is also important to investigate how different
mobility models will affect the performance of caching strategies.}
\item \textit{Privacy issues}: In order to take advantage of the user mobility
pattern, some personal information, e.g., home locations and work
place locations, may be divulged in the collected mobility information.
This will certainly cause some concerns on the privacy issues. Thus,
how to extract the useful user mobility information without touching
the individual privacy is important. Location obfuscation and fake
location injection mechanisms may serve as potential approaches for
anonymous traces.
\end{itemize}
\bibliographystyle{IEEEtran}
\bibliography{IEEEabrv,MAC_ref}

\end{document}